\documentclass[twocolumn]{aa}  

\usepackage{graphicx}
\usepackage{txfonts}
\usepackage{multirow} 
\usepackage{textcomp} 
\usepackage[colorlinks=true,linkcolor=red,citecolor=blue]{hyperref}
\usepackage{ulem}
\usepackage{booktabs} 
\usepackage{xcolor} 
\usepackage{float}

\begin{document} 

\title{Accurate transition and hyperfine data in \ion{Ag}{I} from Multiconfiguration Dirac-Hartree-Fock
and Relativistic Coupled-Cluster methods}

\author{P. J\"onsson\inst{1} \and B.~K. Sahoo\inst{2}
\and S. Caliskan\inst{3} \and A.~M. Amarsi\inst{3}}

\institute{
Department of Materials Science and Applied Mathematics, Malm\"o University, SE-205 06, Malm\"o, Sweden 
\and
Atomic, Molecular and Optical Physics Division, Physical Research Laboratory, Navrangpura, Ahmedabad 380009, India
\and
Theoretical Astrophysics,
Department of Physics and Astronomy,
Uppsala University, Box 516, SE-751 20 Uppsala, Sweden
}

 \abstract
{Silver is a key tracer of the weak r-process in late-type stars.
However, when the assumption of local thermodynamic equilibrium (LTE)
needs to be relaxed, accurate abundance determinations {become even more sensitive to complete sets of reliable transition data.}}
{The aim of this work is to provide accurate and extensive results of excitation energies, {radiative} transition and hyperfine data for \ion{Ag}{I}.}
{The Multiconfiguration Dirac–Hartree–Fock (MCDHF) and relativistic coupled-cluster (RCC) methods were used in the present work. The quantitative and qualitative evaluation (QQE)
approach is applied to the MCDHF transition rates to estimate the
uncertainty according to the National Institute of Science and Technology  Atomic Spectroscopic Data (NIST ASD) terminology.}
{Excitation energies, transition data and hyperfine structure constants were calculated for 18 states up to $4d^{10}8s$. 57 electric dipole (E1) transition rates and weighted oscillator strengths are 
computed and estimated to be in the following NIST ASD uncertainty classes; 4 in AA, 12 in A+, 5 in A, 13 in B+, 6 in B, 4 in C+ 
with AA $\leq 1\%$, A+ $\leq 2\%$, A $\leq 3\%$, B+ $\leq 7\%$, B $\leq 10\%$,
C+ $\leq 18\%$. The remaining transitions, mainly weak transitions involving the $4d^95s^2$ states, are estimated
to be in the E class $>50\%$. The computed lifetimes from both the MCDHF and RCC methods are in good mutual agreement and mostly fall within the error bars
of available experimental values from laser induced fluorescence (LIF) measurements. The $4d^95s^2~^2D_{5/2}$ metastable state, important for establishing the ionization balance, decay through an
E2 transition to the ground state. The calculated lifetime is 163 ms. The computed hyperfine interaction constants from the MCDHF and RCC methods are in good agreement and compare well with the scattered experimental constants. }
{}

\keywords{Atomic data --- Atomic processes --- Line: formation ---  Radiative transfer --- Stars: abundances ---
Methods: numerical}
\titlerunning{Accurate theoretical transition  data in \ion{Ag}{I}}
\authorrunning{P. J\"onsson et al.}

\maketitle

\section{Introduction}
The origins of the elements heavier than iron ($Z=26$), and specifically the
relative yields of different astrophysical sites
of the rapid neutron capture process (r-process), remains
an open question \citep[e.g.][]{2021RvMP...93a5002C}.
There is evidence for several different patterns of r-process yields,
including a weak r-process that may produce
elements up to the second r-process peak
($N=82$; e.g.~\citealt{2023A&ARv..31....1A},
\citealt{2023EPJA...59...12T}).
In this context, silver ($Z=47$) 
is of great interest:
it is predicted to form mostly from the r-process
\citep[e.g.][]{2020MNRAS.491.1832P},
and its lines can be detected in stellar spectra.
Consequently, silver abundances in late-type stars are
key diagnostics for understanding the weak r-process
\citep[e.g.][]{2012A&A...545A..31H,
2015A&A...579A...8W,2025ApJ...991..112H}.

However, the accuracy of stellar abundance determinations is closely coupled
to the input atomic data \citep[e.g.][]{2009PhST..134a4004M,
2025RNAAS...9..123S}.
The transition rate or oscillator strength
has a one-to-one relationship with the
abundance inferred from a given spectral line,
and hyperfine splitting may also significantly impact analyses that are based on saturated lines 
\citep[e.g.][]{2018ApJ...866...52T}.
Furthermore, when the assumption of local thermodynamic equilibrium
(LTE) is relaxed, 
{radiative transition data are required not only for the diagnostic lines used for abundance determinations, but also for all of the lines that couple levels included in the model atom.  This is because radiative rates enter the coupled statistical-equilibrium equations for all levels and can therefore affect inferred abundances indirectly through their influence on the non-LTE level populations. Therefore,} a comprehensive and accurate set of transition rates is essential for non-LTE diagnostics \citep[e.g.][]{2024ARA&A..62..475L}.

In the case of silver \citep[e.g.][]{2023ExA....55..133H}, abundances are usually inferred
from the \ion{Ag}{I} $328\,\mathrm{nm}$ and
$338\,\mathrm{nm}$ resonance lines
($5s\,{^2}S_{1/2}\rightarrow
5p\,{^2}P^{\mathrm{o}}_{3/2,1/2}$).
Although the transition rates from the NIST database \citep{NISTdata} (given with 10$\%$ accuracy) and the measured hyperfine structure data \citep{hansen_silver_2012} for these two lines are well constrained, a detailed analysis of the solar spectrum based on
three-dimensional radiation-hydrodynamics
\citep{10015TP} 
suggests that these lines are prone to large non-LTE effects, at least in the Sun \citep{2021A&A...653A.141A},
and thus may be indirectly
affected by the full set of transition data.
In order to test this and ultimately derive
accurate silver abundances in late-type stars,
a model atom for \ion{Ag}{I} is under construction
(Caliskan et al.~in prep.), which in turn requires a reliable and as complete as possible set of transition data for \ion{Ag}{I}.

Being a nominal one-electron system, Ag I has been extensively studied with respect to its hyperfine structures, transition rates, and lifetimes using a variety of experimental and theoretical methods.
The lifetimes of the first two excited states, $5p_{1/2,3/2}$, were measured with high precision by \cite{5944TP} using a picosecond delayed-coincidence technique. The reported uncertainties are as low as 0.5\,\%, making these lifetimes among the most accurately known in atomic physics and highly valuable for benchmarking purposes, see also \cite{Carlsson_1989} for an in-depth discussion about the error sources. Hyperfine interaction constants were also determined through quantum beat spectroscopy.
Lifetimes of states in the $s$ and $d$ sequences, up to $10s_{1/2}$ and $9d_{3/2}$, were measured using laser-induced fluorescence (LIF) combined with two-step excitations \citep{5681TP}. These results showed good agreement with multiconfiguration Hartree–Fock (MCHF) calculations based on a model potential, as reported in the same study. Similarly, the lifetimes of the $6p_{1/2,3/2}$ and $7p_{1/2,3/2}$ states were measured using LIF techniques \citep{5647TP,5835TP}, and these works also provide hyperfine interaction constants.
The hyperfine interaction constant of the $4d^95s^2~^2D_{5/2}$ state was measured with exceptional accuracy by \cite{PhysRev.150.59} using the atomic-beam magnetic-resonance method.

On the theoretical side, numerous studies have been conducted, as reflected in the NIST database \citep{NISTdata}.
Here, we highlight those most relevant to our work. The pseudo-relativistic Hartree–Fock (HFR) method was employed by \cite{10654TP} to compute atomic structures and radiative parameters for highly excited states in the $s$, $p$, $d$ and $f$ sequences, including configurations with holes in the $4d$-subshell. The weakest bound electron potential model (WBEPM) was used by \cite{7167TP} to study transition probabilities, while \cite{8781TP} applied the Fues model potential to calculate transition dipole matrix elements. The latter work is particularly comprehensive, providing extensive tables comparing oscillator strengths from both experimental and theoretical sources.
Third-order relativistic many-body theory was utilized by \cite{PhysRevA.68.062505} to compute excitation energies and lifetimes of the lowest states with $s,p,d,f$, and $g$ symmetries. More recently, \cite{10.1063/5.0128225} applied Fock-space multi-reference coupled cluster (FSMRCC) theory to determine excitation energies, hyperfine constants, and transition parameters for the $5s_{1/2}$ and $5p_{1/2,3/2}$ states. Additionally, hyperfine constants for these states were calculated using the multiconfiguration Dirac–Hartree–Fock (MCDHF) method by \cite{Song_2007}.

In this work, we employ the advanced  FSMRCC theory together with the combined multiconfiguration Dirac–Hartree–Fock and relativistic configuration interaction (MCDHF/RCI) methods to compute excitation energies, hyperfine interaction constants, transition data, and lifetimes for the $5s_{1/2},6s_{1/2},7s_{1/2},8s_{1/2},5d_{3/2,5/2},6d_{3/2,5/2},4d^95s^2~^2D_{3/2,5/2}$ even and the
$5p_{1/2,3/2},6p_{1/2,3/2},7p_{1/2,3/2}$, $4f_{5/2,7/2}$ odd states in Ag I. 

\section{Theory and procedure}

 Detailed descriptions of the MCDHF/RCI methods  can be found in \cite{ grant2007relativistic, atoms11010007}, and in the recent comparative study with the FSMRCC method by \cite{bpg7-wm48}. Therefore, only the computational procedures are outlined here. The FSMRCC method employed here is similar that is applied by \cite{10.1063/5.0128225}, but it incorporates triple excitations and Breit interactions in the calculations.
 
\subsection{MCDHF and RCI}

All MCDHF and RCI calculations in this work were performed using the GRASPG package \citep{SI2025109604}, an extension of GRASP \citep{FROESEFISCHER2019184} based on configuration state function generators (CSFGs), enabling significantly faster computations with reduced resource requirements \citep{LI2023108562}. 
An initial MCDHF calculation was carried out for states belonging to the reference configurations \(4d^{10}\{5s,6s,7s,8s,5p,6p,7p,5d,6d,4f\}\) with additional states from \(4d^{10}\{9s,10s,8p,9p,7d\}\) to introduce spatially extended orbitals into the basis. This is essential for achieving agreement between transition parameters in the length and velocity gauges and reducing uncertainties \citep{9990TP,atoms7040106}. 
Subsequent MCDHF calculations employed CSF expansions generated by allowing single and double (SD) excitations to progressively larger orbital sets from subshells of the reference configurations down to \(4s^24p^64d^{10}\), with at most one excitation from this layer. The closed core \(1s^22s^22p^63s^23p^63d^{10}\) remained inactive. Following the terminology of \cite{MCHF}, these calculations account for radial and core-valence (CV) correlation. Orbital sets were increased layer by layer as summarized in Table~\ref{tab:orbital}. 
Finally, RCI calculations were performed using the same CSF expansions, now including Breit and QED effects.

\subsection{FSMRCC}  

Computational details of our FSMRCC method are discussed in a series of previous works \citep{bijaya0,bijaya1,bijaya2}. We have considered Gaussian type orbitals (GTOs) to construct single particle orbitals using the Dirac-Hartree-Fock (DHF) method \citep{GTOsbasis}. We generate orbitals up to $i$-symmetry considering 40 GTOs for each angular momentum symmetry. The Breit interaction terms are included self-consistently at the DHF and FSMRCC methods. Convergence in the results are verified by considering singles and doubles excitations in the calculations first, then by including triple excitations in the FSMRCC method \citep{bijaya3,bijaya03}. In the estimations of the magnetic dipole structure constants ($A_{\text{hfs}}$), we have also considered corrections due to the Bohr-Weisskopf (BW) effects. Detailed procedure to estimate the BW effects are discussed in \citet{bijaya3} and \citet{bijaya4}.  Results with singles and doubles approximation in the FSMRCC method are given under RCCSD method and results from the single, doubles and triples approximated FSMRCC method are listed under RCCSDT method.

\begin{table}[h]
\begin{center}
  \caption{Layers of correlation orbitals, in non-relativistic notation, used in the MCDHF/RCI calculations.}
    \begin{tabular}{cl} \hline \hline
     layers & orbitals \\ \hline
     layer 1   &  $\{10s~9p~~~7d~~\,5f~~\,5g\}$\\
     layer 2   &  $\{11s~10p~~8d~~\,6f~~\,6g\}$\\
     layer 3   &  $\{12s~11p~~9d~~\,7f~~\,7g\}$\\
     layer 4   &  $\{13s~12p\,~10d~~8f~~8g\}$\\
     layer 5   &  $\{14s~13p\,~11d~~9f~~9g\}$\\
     layer 6   &  $\{15s~14p\,~12d~~10f~10g\}$\\ \hline 
    \end{tabular}
    \label{tab:orbital}
\end{center}
\end{table}

\section{Results and discussion}
\subsection{Excitation energies}
The excitation energies from the MCDHF/RCI calculations are presented in Table~\ref{tab:energyCV} as functions of the increasing layers of correlation orbitals. The state \(4d^{10}7s~^2S_{1/2}\) initially appears too high but approaches experimental values after two correlation layers. In contrast, the excitation energies of {the \(4d^{9}5s^2~^2D\)} states remain significantly overestimated in the CV model by about 9000~cm\(^{-1}\). To ensure accurate transition parameters, these energies were fine-tuned (FT) by manually adjusting the dominant Hamiltonian matrix element \citep{atoms11040070}, chosen to reproduce correct transition energies to the \(^{2}P\) and \(^{2}F\) states. 
On average, excitation energies in the CV correlation model are overestimated by 900~cm\(^{-1}\) for lower states and up to 1800~cm\(^{-1}\) for higher states. Including core-core (CC) correlations through double substitutions from the core, as well as higher-order effects from triple and quadruple (TQ) substitutions, improves agreement with experiment at the cost of substantially larger CSF expansions.

   \begin{table*}
   \begin{center}
  \caption{Calculated excitation energies (in cm$^{-1}$) at different layers using the MCDHF/RCI method. The column ``FT" reports the energies for which the diagonal Hamiltonian matrix 
  elements have been fine-tuned to give accurate transition energies to the $^2P$ and $^2F$ states. The excitation energies are compared with the experimental values from the NIST ASD database \citep{NISTdata}.
NCSFs is the total number of CSFs in the wave function expansion.}
     \begin{tabular}{lrrrrrrrr} \hline \hline
      State                    & layer 1  & layer 2  & layer 3  & layer 4  & layer 5  & layer 6  & FT    &  Experiment \\ \hline
      $4d^{10}5s~^2S_{1/2}$    &          &          &          &          &        0 & 0        &       &      0 \\
      $4d^{10}5p~^2P_{1/2}$    & 29666.09 & 30354.06 & 30471.03 & 30566.38 & 30582.32 & 30603.42 & 30603.42 & 29~552.057 \\
      $4d^{10}5p~^2P_{3/2}$    & 30566.34 & 31311.67 & 31443.87 & 31544.60 & 31563.46 & 31584.35 & 31584.35 & 30~472.665 \\
      $4d^{9}5s^2~^2D_{5/2}$   & 37669.11 & 38789.07 & 39066.71 & 39222.35 & 39260.77 & 39305.78 & 31323.21 & 30~242.298 \\
      $4d^{9}5s^2~^2D_{3/2}$   & 42425.88 & 43546.51 & 43822.39 & 43978.36 & 44016.75 & 44061.73 & 35874.87 & 34~714.226 \\
      $4d^{10}6s~^2S_{1/2}$    & 42626.16 & 43743.93 & 43959.01 & 44094.63 & 44105.87 & 44127.67 & 44127.67 & 42~556.147\\
      $4d^{10}6p~^2P_{1/2}$    & 48319.60 & 49518.92 & 49757.99 & 49909.21 & 49941.78 & 49969.91 & 49969.91 & 48~297.406  \\
      $4d^{10}6p~^2P_{3/2}$    & 48519.77 & 49726.64 & 49969.27 & 50121.69 & 50153.99 & 50182.13 & 50182.13 & 48~500.810 \\
      $4d^{10}5d~^2D_{3/2}$    & 48769.42 & 49978.98 & 50236.71 & 50400.94 & 50425.84 & 50453.33 & 50453.33 & 48~743.969\\
      $4d^{10}5d~^2D_{5/2}$    & 48800.14 & 49997.10 & 50253.26 & 50417.21 & 50442.01 & 50471.14 & 50471.14 & 48~764.219\\
      $4d^{10}7s~^2S_{1/2}$    & 55498.52 & 56793.66 & 53384.75 & 53548.46 & 53580.45 & 53604.78 & 53604.78 & 51~886.965\\
      $4d^{10}7p~^2P_{1/2}$    & 54025.44 & 55316.33 & 55582.43 & 55740.77 & 55773.98 & 55802.99 & 55802.99 & 54~041.037 \\
      $4d^{10}7p~^2P_{3/2}$    & 54104.65 & 55397.83 & 55665.92 & 55825.00 & 55856.35 & 55885.40 & 55885.40 & 54~121.108\\
      $4d^{10}6d~^2D_{3/2}$    & 54163.05 & 55455.75 & 55734.23 & 55903.50 & 55929.97 & 55960.34 & 55960.34 & 54~203.119 \\
      $4d^{10}4f~^2F_{5/2}$    & 54105.56 & 55456.44 & 55759.78 & 55938.03 & 55938.98 & 56002.28 & 56002.28 & 54~204.729\\
      $4d^{10}4f~^2F_{7/2}$    & 54105.49 & 55456.37 & 55759.71 & 55937.96 & 55968.58 & 56002.35 & 56002.35 & 54~204.745 \\
      $4d^{10}6d~^2D_{5/2}$    & 54178.48 & 55465.47 & 55743.41 & 55912.57 & 55968.65 & 55969.55 & 55969.55 & 54~213.564\\
      $4d^{10}8s~^2S_{1/2}$    & 57347.35 & 58674.23 & 57144.94 & 57309.58 & 57317.80 & 57342.71 & 57342.71 & 55~581.246 \\ \hline
      NCSF                     & 13867    &  23803   &  36501   &  51961    & 70183    & 91167    & 91167    &            \\\hline 
     \end{tabular}
     \label{tab:energyCV}
     \end{center}
 \end{table*}

 The second ionization potential of the ground state and excitation energies of other states from the FSMRCC method are presented in Table~\ref{tab_execc}. As shown, the excitation energies increase progressively from DHF to RCCSD to RCCSDT, indicating that electron correlation contributes more significantly to the ground state than to the excited states. Breit corrections are relatively small, and the final values are taken as RCCSDT results with added Breit contributions. Comparison with experimental data \citep{NISTdata} shows good agreement, except for the {\(4d^95s^2~^2D\)} states, which remain overestimated by about 4800~cm\(^{-1}\). For a general discussion about the computational difficulties for 
 core excited states such as {\(4d^95s^2~^2D\)}, see for example \citep{Caliskan_2024}.
 
\begin{table*}
\begin{center}
\caption{Calculated second ionization potential and excitation energies (in cm$^{-1}$) at different levels of approximation in the FSMRCC theory and comparison with the experimental values from the NIST ASD database \citep{NISTdata}.}
\begin{tabular}{c|ccc c| cc} \hline \hline
State  & DHF  &  RCCSD  & RCCSDT & Breit & Final & Experiment \\ \hline
\multicolumn{7}{c}{Second ionization potential} \\ 
$4d^{10}5s~^2S_{1/2}$    &  50376.11  &  60419.22  &  61455.66 & $-58.71$ & $61396.95$ &  61106.45 \\       
\hline \\ 
\multicolumn{7}{c}{Excitation energies} \\ 
$4d^{10}5p~^2P_{1/2}$  & 23645.86  & 29405.57  & 29817.80  &  $-23.15$ & 29794.65 & 29~552.057 \\
$4d^{10}5p~^2P_{3/2}$  & 24227.72  & 30324.82  & 30774.22  & $-35.20$   & 30739.02  & 30~472.665 \\
$4d^{9}5s^2~^2D_{5/2}$ & 61491.97  & 34897.95  & --  &  198.10 & 35096.05 & 30~242.298 \\
$4d^{9}5s^2~^2D_{3/2}$ & 66890.45  & 39449.04  & --  & 46.05 & 39495.09 & 34~714.226 \\
$4d^{10}6s~^2S_{1/2}$  & 33261.14  & 41963.06  & 42897.62  & $-50.64$ & 42846.98 & 42~556.147\\
$4d^{10}6p~^2P_{1/2}$  & 38590.28  & 47737.47  & 48666.10  & $-50.83$  & 48615.27  & 48~297.406  \\
$4d^{10}6p~^2P_{3/2}$  & 38758.13  & 47950.68  & 48879.08  & $-53.48$ & 48825.60  & 48~500.810 \\
$4d^{10}5d~^2D_{3/2}$ & 38393.71  & 48101.67  & 49088.29  & $-58.05$  & 49030.24  & 48~743.969\\
$4d^{10}5d~^2D_{5/2}$  & 38409.24  & 48122.18  & 49113.99  & $-58.67$ & 49055.32 & 48~764.219\\
$4d^{10}7s~^2S_{1/2}$  & 41637.14  & 51231.60  & 52235.83 & $-55.99$  & 52179.84 & 51~886.965\\
$4d^{10}7p~^2P_{1/2}$  & 43717.54  & 53410.22  & 54405.47  & $-55.56$  & 54349.91 & 54~041.037 \\
$4d^{10}7p~^2P_{3/2}$  & 43788.97  & 53496.26  & 54490.84  & $-56.61$ & 54434.23 & 54~121.108\\
$4d^{10}6d~^2D_{3/2}$  & 43633.55  & 53535.84  & 54555.13 &  $-58.40$ & 54496.73 & 54~203.119 \\
$4d^{10}4f~^2F_{5/2}$  & 43518.71  & 53525.56  & 54558.28  & $-58.67$  & 54499.61  & 54~204.729\\
$4d^{10}4f~^2F_{7/2}$  & 43518.64  & 53525.48  &  54558.04 & $-58.67$  & 54499.37  & 54~204.745 \\
$4d^{10}6d~^2D_{5/2}$  & 43641.70 & 53545.90  & 54563.71  & $-58.70$  & 54505.01 & 54~213.564\\
$4d^{10}8s~^2S_{1/2}$  & 45071.25  & 54908.80 &  55931.10 &  $-57.46$ & 55873.64  & 55~581.246 \\ 
\hline 
\end{tabular}
\label{tab_execc}
\end{center}
\end{table*}

By comparing the MCDHF/RCI results with those from the RCCSD and RCCSDT calculations, we find that the FSMRCC values are generally closer to experiment. Furthermore, unlike MCDHF/RCI, the FSMRCC calculations correctly reproduce the ordering of the closely spaced \(4d^{10}6d\) and \(4d^{10}4f\) configurations.

\subsection{Hyperfine interaction constants}
Silver has two stable isotopes, \(^{107}\mathrm{Ag}\) and \(^{109}\mathrm{Ag}\), with natural abundances of 51.839\,\% and 48.161\,\%, respectively. The dominant isotope, \(^{107}\mathrm{Ag}\), has nuclear spin \(I = 1/2\) and a magnetic dipole moment \(\mu = -0.113570\,\mu_N\), while \(^{109}\mathrm{Ag}\) has \(I = 1/2\) and \(\mu = -0.1306905\,\mu_N\). 
Due to the interaction between the electrons and the magnetic dipole moment, the recorded spectral lines are split into hyperfine components, with splittings governed by the magnetic dipole interaction constants 
\(A_{\mbox{\scriptsize hfs}}\) of the upper and lower states \citep{MCHF}.

The hyperfine interaction constants \(A_{\mbox{\scriptsize hfs}}\) from the MCDHF/RCI method are presented in Table~\ref{tab:hfs_MCDHF} as functions of the correlation layers. The convergence pattern shows a tendency for the absolute values of the constants in the \(s\) and \(p\) sequences to increase as more core-valence (CV) correlation is included, with particularly large changes for the \(7s\) and \(8s\) states. This behavior is consistent with general trends observed for nominal one-electron systems \citep{PhysRevA.101.062510}. 
Calculations accounting for CV correlation tend to overshoot the absolute values for the lowest states, which agrees with observations for similar systems where the inclusion of configuration state functions (CSFs) from double (D) substitutions, also accounting for core-core (CC) correlation, lowers the values and improves agreement with experiment \citep{PhysRevA.101.062510}. Due to the resulting large expansion sizes, these computations have not been attempted.

  \begin{table*}
  \begin{center}
    \caption{Calculated  hyperfine interaction constants \(A_{\mbox{\scriptsize hfs}}\)  (in MHz) of $^{107}$Ag  at different layers using the MCDHF/RCI method and comparison with the experimental values.}
     \begin{tabular}{lrrrrrrr} \hline \hline
      State                    &  layer 1         &  layer 2        &  layer 3      &  layer 4    &  layer 5   &  layer 6 &  Experiment \\ \hline
      $4d^{10}5s~^2S_{1/2}$    & $-1676$          & $-1736$         & $-1765$       & $-1847$     & $-1847$    & $-1845$  &   $-1712.512111(18)^{(a)}$\\
      $4d^{10}5p~^2P_{1/2}$    & $-164$           & $-175$          & $-186$        & $-192$      & $-192$     & $-193$   &   $-175.4(17)^{(b)}$ \\
      $4d^{10}5p~^2P_{3/2}$    & $-27.2$           & $-31.1$         & $-33.7$       & $-34.8$     & $-34.4$   & $-34.3$  &   $-31.7(5)^{(b)}$\\
      $4d^{9}5s^2~^2D_{5/2}$   & $-145$           & $-148$          & $-148$        & $-147$      & $-146$    & $-145$   &   $-126.2818(1)^{(c)}$\\
      $4d^{9}5s^2~^2D_{3/2}$   & $-267$            & $-266$          & $-265$        & $-266$      & $-266$    & $-268$   &               \\
      $4d^{10}6s~^2S_{1/2}$    & $-199$            & $-204$          & $-219$      & $-239$    & $-241$     & $-241$   &               \\
      $4d^{10}6p~^2P_{1/2}$    & $-32.9$           & $-34.0$         & $-39.4$       & $-39.9$     & $-40.6$   & $-40.3$  &   $-38.7(10)^{(d)}$ \\
      $4d^{10}6p~^2P_{3/2}$    & $-6.7$            & $-7.2$          & $-8.5$        & $-8.5$      & $-8.6$    & $-8.5$   &   $-9.05(25)^{(d)}$ \\
      $4d^{10}5d~^2D_{3/2}$    & $-1.4$            & $-1.4$          & $-1.7$        & $-1.8$      & $-1.8$    & $-1.7$   &    \\
      $4d^{10}5d~^2D_{5/2}$    & $-0.8$            & $-0.8$          & $-0.9$        & $-0.9$      & $-0.9$    & $-0.9$   &    \\
      $4d^{10}7s~^2S_{1/2}$    & $-36.5$           & $-40.4$         & $-79.3$       & $-81.4$   & $-81.5$    & $-81.6$  &    \\
      $4d^{10}7p~^2P_{1/2}$    & $-11.5$           & $-11.8$         & $-15.0$       & $-15.0$     & $-15.7$   & $-15.4$  &    \\
      $4d^{10}7p~^2P_{3/2}$    & $-2.6$            & $-2.8$          & $-3.7$        & $-3.6$      & $-3.8$    & $-3.7$   &   $-4.5(2)^{(e)}$ \\
      $4d^{10}6d~^2D_{3/2}$    & $-0.6$            & $-0.6$          & $-0.8$        & $-0.8$      & $-0.8$    & $-0.8$   &       \\
      $4d^{10}4f~^2F_{5/2}$    & $-0.01$           & $-0.01$         & $-0.01$       & $-0.01$     & $-0.01$   & $-0.01$   &   \\
      $4d^{10}4f~^2F_{7/2}$    & $-0.009$          & $-0.009$        & $-0.009$      & $-0.009$    & $-0.009$  & $-0.009$ &  \\
      $4d^{10}6d~^2D_{5/2}$    & $-0.4$            & $-0.4$          & $-0.4$        & $-0.4$      & $-0.4$    & $-0.4$   &   \\
      $4d^{10}8s~^2S_{1/2}$    & $-15.4$           & $-16.2$         & $-26.1$       & $-30.9$     & $-36.5$    & $-37.0$  &  \\ \hline 
     \end{tabular}
     \tablefoot{
     \tablefoottext{a}{\cite{Agground}};
     \tablefoottext{b}{\cite{5944TP}};
     \tablefoottext{c}{\cite{PhysRev.150.59}};
     \tablefoottext{d}{\cite{5647TP}};
     \tablefoottext{e}{\cite{5835TP}}
     }
     \label{tab:hfs_MCDHF}
     \end{center}
 \end{table*}

The FSMRCC values for the hyperfine interaction constants \(A_{\mathrm{hfs}}\) at different levels of approximation are given in Table~\ref{tab_hfs}. We have also included the BW corrections \citep{Roberts2022BohrWeisskopf} in the estimation of these quantities using the RCCSD method. The table shows that the Breit and BW corrections are generally small, except for the ground state where they are more significant. 

\begin{table*}
\begin{center}
\caption{Calculated  hyperfine interaction constants \(A_{\mbox{\scriptsize hfs}}\)  (in MHz) of $^{107}$Ag at different levels of approximation in the FSMRCC theory and comparison with the experimental values.}
\begin{tabular}{c|rrrrr | rr} \hline \hline
State       & DHF     &   RCCSD  & RCCSDT  & BW & Breit &  Final &  Experiment \\ \hline
$4d^{10}5s~^2S_{1/2}$    & $-1198.13$ & $-1761.06$ & $-1799.57$ & 8.78 & $-0.01$ & $-1790.80$ & $-1712.512111(18)^{(a)}$ \\
$4d^{10}5p~^2P_{1/2}$    & $-104.46$ & $-173.89$ & $-181.96$ & 0.06 & $0.52$ & $-181.38$ & $-175.4(17)^{(b)}$ \\
$4d^{10}5p~^2P_{3/2}$    & $-16.06$ & $-29.31$ & $-32.07$ & 0.01 & $0.01$ & $-32.05$ & $-31.7(5)^{(b)}$ \\
$4d^{9}5s^2~^2D_{5/2}$   & $-116.78$ & $-194.35$ &  --  & $-0.02$ & $-0.96$ & $-195.33$ & $-126.2818(1)^{(c)}$\\
$4d^{9}5s^2~^2D_{3/2}$   & $-287.87$ & $-479.59$ &  --   & 0.02 & $-1.87$ & $-481.44$ & \\
$4d^{10}6s~^2S_{1/2}$    & $-195.20$ & $-239.90$ & $-238.96$ & 1.20 & $-0.06$ & $-237.82$ & \\
$4d^{10}6p~^2P_{1/2}$    & $-29.02$ & $-39.27$ & $-39.67$ & 0.02 & 0.05 & $-39.60$ & $-38.7(10)^{(d)}$ \\
$4d^{10}6p~^2P_{3/2}$    & $-0.23$ & $-7.00$ & $-7.78$ & $\sim 0.0$ & $\sim 0.0$ & $-7.78$  & $-9.05(25)^{(d)}$ \\
$4d^{10}5d~^2D_{3/2}$    & $-0.73$ & $-1.59$ & $-1.79$ &  $\sim 0.0$ & $-0.01$ & $-1.80$ & \\
$4d^{10}5d~^2D_{5/2}$    & $-0.32$ & $-0.64$ & $-65$ &  $\sim 0.0$ & $\sim 0.0$ & $-0.65$ &\\
$4d^{10}7s~^2S_{1/2}$    & $-69.50$ & $-82.04$ & $-81.43$ &  0.41 & $-0.10$ & $-81.12$ & \\
$4d^{10}7p~^2P_{1/2}$    & $-12.21$ & $-15.77$ & $-15.80$ &  0.01 & $-0.04$ & $-15.83$ & \\
$4d^{10}7p~^2P_{3/2}$    & $-1.94$ & $-2.87$ & $-3.27$ &  $\sim 0.0$ & $\sim 0.0$ & $-3.27$ & $-4.5(2)^{(e)}$ \\
$4d^{10}6d~^2D_{3/2}$  & $-0.38$ & $-0.82$ & $-0.85$ &  $\sim 0.0$ & $\sim 0.0$ & $-0.85$ & \\
$4d^{10}6d~^2D_{5/2}$    & $-0.17$ & $-0.31$ & $-0.33$ &   $\sim 0.0$ & $\sim 0.0$ & $-0.33$ & \\
$4d^{10}4f~^2F_{5/2}$  & $-0.011$ & $-0.012$ & $-0.014$  &  $\sim 0.0$ & $\sim 0.0$ & $-0.014$ & \\
$4d^{10}4f~^2F_{7/2}$  & $-0.006$ & $-0.007$ & $-0.008$ &  $\sim 0.0$ & $\sim 0.0$ & $-0.008$ & \\
$4d^{10}8s~^2S_{1/2}$    & $-32.66$ & $-37.88$ & $-37.51$ & 0.19 & $-0.01$ & $-37.33$ & \\ \hline 
\end{tabular}

     \tablefoot{
     \tablefoottext{a}{\cite{Agground}};
     \tablefoottext{b}{\cite{5944TP}};
     \tablefoottext{c}{\cite{PhysRev.150.59}};
     \tablefoottext{d}{\cite{5647TP}};
     \tablefoottext{e}{\cite{5835TP}}
     }
\label{tab_hfs}
\end{center}
\end{table*}

For convenience, the hyperfine interaction constants from both the MCDHF/RCI and FSMRCC methods are summarized in Table~\ref{tab:hfs_all}, along with other theoretical results and experimental data. Comparing the results, we find that the FSMRCC values agree better with experiment for the lower states than the MCDHF/RCI results, although the overall agreement is good. For the {\(4d^95s^2~^2D_{5/2}\)} state, where the experimental value is known with high precision, the MCDHF/RCI calculation provides a better estimate. The MCDHF calculations by \cite{Song_2007} for the ground and first excited states remain the most accurate to date.

  \begin{table*}
  \begin{center}
    \caption{Hyperfine interaction constants \(A_{\mbox{\scriptsize hfs}}\)  (in MHz) of $^{107}$Ag. Comparison between different calculated values and experiment.}
     \begin{tabular}{lrrrrr} \hline \hline
      State                    &  MCDHF/RCI & FSMRCC        & MCDHF$^{(a)}$ & FSMRCC$^{(b)}$ & Experiment \\ \hline
      $4d^{10}5s~^2S_{1/2}$    & $-1845$    & $-1790.80$ & $-1724$ & $-1726$ & $-1712.512111(18)^{(c)}$\\
      $4d^{10}5p~^2P_{1/2}$    & $-193$     & $-181.38$  & $-180$  & $-167$  & $-175.4(17)^{(d)}$ \\
      $4d^{10}5p~^2P_{3/2}$    & $-34.3$    & $-32.05$   & $-30$   & $-28$ & $-31.7(5)^{(d)}$\\
      $4d^{9}5s^2~^2D_{5/2}$   & $-145$     & $-195.33$  &         & & $-126.2818(1)^{(e)}$\\
      $4d^{9}5s^2~^2D_{3/2}$   & $-268$     & $-481.44$  &         & &             \\
      $4d^{10}6s~^2S_{1/2}$    & $-241$     & $-237.82$  &         & &             \\
      $4d^{10}6p~^2P_{1/2}$    & $-40.3$    & $-39.60$   &         & & $-38.7(10)^{(f)}$ \\
      $4d^{10}6p~^2P_{3/2}$    & $-8.5$     & $-7.78$    &         & & $-9.05(25)^{(f)}$ \\
      $4d^{10}5d~^2D_{3/2}$    & $-1.7$     & $-1.80$    & & &  \\
      $4d^{10}5d~^2D_{5/2}$    & $-0.9$     & $-0.65$    & & &  \\
      $4d^{10}7s~^2S_{1/2}$    & $-81.6$    & $-81.12$   & & &  \\
      $4d^{10}7p~^2P_{1/2}$    & $-15.4$    & $-15.83$   & & &  \\
      $4d^{10}7p~^2P_{3/2}$    & $-3.7$     & $-3.27$    & & & $-4.5(2)^{(g)}$ \\
      $4d^{10}6d~^2D_{3/2}$    & $-0.8$     & $-0.85$    & & &     \\
      $4d^{10}4f~^2F_{5/2}$    & $-0.01$    & $-0.014$   & & & \\
      $4d^{10}4f~^2F_{7/2}$    & $-0.009$   & $-0.008$   & & & \\
      $4d^{10}6d~^2D_{5/2}$    & $-0.4$     & $-0.33$    & & & \\
      $4d^{10}8s~^2S_{1/2}$    & $-37.0$    & $-37.33$   & & & \\ \hline 
     \end{tabular}

\tablefoot{
\tablefoottext{a}{\cite{Song_2007}};
\tablefoottext{b}{\cite{10.1063/5.0128225}};
\tablefoottext{c}{\cite{Agground}};
\tablefoottext{d}{\cite{5944TP}};
\tablefoottext{e}{\cite{PhysRev.150.59}};
\tablefoottext{f}{\cite{5647TP}};
\tablefoottext{g}{\cite{5835TP}}}
     \label{tab:hfs_all}

     \end{center}
 \end{table*}

\subsection{Transition rates and oscillator strengths}
The fundamental quantity from which all parameters for electric dipole (E1) transitions can be derived is the reduced matrix element \citep{bpg7-wm48}. Table~\ref{tab_E1} compares the reduced matrix elements obtained at different levels of approximation in the FSMRCC method (using the length gauge) with those from the MCDHF/RCI calculations. The median relative difference between the two methods is 1.7\%. While most transitions show very small differences, notable exceptions occur for the \(6p_{1/2,3/2} \rightarrow 5s_{1/2}\) and \(7p_{1/2,3/2} \rightarrow 5s_{1/2}\) transitions. As discussed in \citet{atoms7040106}, the length-gauge matrix elements in MCDHF/RCI for transitions from highly excited states to low-lying states are sensitive to the completeness of the radial orbital basis, which affects the outer part of the wave function. 
In contrast, the velocity-gauge matrix elements are less sensitive to basis incompleteness and provide more accurate values. Indeed, the velocity-gauge results for these transitions (0.162 and 0.310 for \(6p_{1/2,3/2} \rightarrow 5s_{1/2}\), and 0.0411 and 0.104 for \(7p_{1/2,3/2} \rightarrow 5s_{1/2}\)) are much closer to the FSMRCC values, which do not suffer from this limitation. These transitions also exhibit the largest changes when moving from the DHF approximation to RCCSD and further to RCCSDT.

The transition rates \(A\) from the MCDHF/RCI and FSMRCC calculations are presented in Table~\ref{tab:transrate}. Due to limitations of the FSMRCC approach, transition rates involving the core-excited {\(4d^95s^2~^2D_{3/2,5/2}\)} states could not be computed; these transitions are, however, provided by the MCDHF/RCI calculations. For MCDHF/RCI, the quantitative and qualitative evaluation (QQE) approach of Gaigalas and co-workers \citep{Gaigalas.2022.p281,Kitoviene.2024.p} was used to estimate uncertainties following the NIST ASD \citep{Kramida.2023.p} classification: AA (\(\leq 1\%\)), A\({+}\) (\(\leq 2\%\)), A (\(\leq 3\%\)), B\({+}\) (\(\leq 7\%\)), B (\(\leq 10\%\)), C\({+}\) (\(\leq 18\%\)), C (\(\leq 25\%\)), D\({+}\) (\(\leq 40\%\)), D (\(\leq 50\%\)), and E (\(>50\%\)). 
Overall, there is good agreement between FSMRCC and MCDHF/RCI transition rates, except for the \(6p_{1/2,3/2} \rightarrow 5s_{1/2}\) and \(7p_{1/2,3/2} \rightarrow 5s_{1/2}\) transitions, as discussed above. For most transitions between Rydberg states, uncertainties are within 10\% according to the QQE approach. Transitions connecting the {\(4d^95s^2~^2D_{3/2,5/2}\)} core-excited states and the ordinary \(4d^{10}nl\) Ryd\-berg states are so-called two-electron one-photon (TEOP) transitions. These transitions vanish in the Dirac–Fock approximation and appear only through electron correlation effects, making them both weak and computationally challenging. Consequently, they fall into the E class (uncertainty \(>50\%\)). To the best of our knowledge, these transition rates have not been reported in any previous calculations.
The transition rates for the  \(6p_{1/2,3/2} \rightarrow 5s_{1/2}\) transitions were recently measured using  optical emission spectroscopy (OES) by \cite{ALHIJRY2020106922}
giving the values  $A = (5.53 \pm 1.72) \times 10^5$ s$^{-1}$ and $A = (3.0 \pm 0.91) \times 10^5$ s$^{-1}$ in stark contrast to the present values as well as values 
from the Kurucz database. Furthermore, the OES values contradict the measured lifetimes of the \(6p_{1/2,3/2}\) states \citep{5681TP}.

 The weighted oscillator strengths, \(gf\), from the MCDHF/RCI and FSMRCC calculations are presented in Table~\ref{tab:rategf}. For MCDHF/RCI, the uncertainty estimates from Table~\ref{tab:transrate} have been included. The table also lists values reported by \citet{8781TP} based on the Fues model potential for comparison. Except for the \(6p_{1/2,3/2} \rightarrow 5s_{1/2}\) and \(7p_{1/2,3/2} \rightarrow 5s_{1/2}\) transitions, there is excellent agreement between MCDHF/RCI and FSMRCC, with a mean relative difference of $3.5\, \%$. The Fues model potential values, despite the simplicity of the method, show relatively good agreement with the results from full many-body theories. However, the differences are especially large for the \(6p_{1/2,3/2} \rightarrow 5s_{1/2}\) and \(7p_{1/2,3/2} \rightarrow 5s_{1/2}\) transitions. Indeed, Table~III of \citet{8781TP}, which compiles \(gf\) values from numerous theories, reveals a scatter spanning orders of magnitude for these transitions.

 \subsection{Lifetimes}
Silver is relatively easily accessible for laser excitation, and lifetimes have been measured for states in the \(s\), \(p\), and \(d\) sequences using laser-induced fluorescence (LIF). In Table~\ref{tab:lifetime}, the lifetimes obtained with the MCDHF/RCI and FSMRCC methods are compared with results from other theories as well as with experiment. Both MCDHF/RCI and FSMRCC slightly underestimate the lifetime for the resonance transition, which is very accurately known \citep{5944TP}. A similar trend, though further from experiment, is observed for FSMRCC and CI-MBPT theories \citep{10.1063/5.0128225,10287TP}. 
For the excited states in the \(p\) sequence, the MCDHF/RCI method predicts somewhat longer lifetimes than FSMRCC, in agreement with experiment within the error bars. For the states in the \(d\) and \(f\) sequences, the two methods give lifetimes in very close agreement, which also match experimental values within uncertainties. For the \(s\) sequence, the opposite trend is observed: FSMRCC predicts slightly longer lifetimes than MCDHF/RCI, again in good agreement with experiment. Overall, the agreement between theory and experiment is very good.
The {\(4d^{9}5s^2~^2D_{3/2,5/2}\)} states are metastable and decay only via higher-order magnetic and electric transitions.
The lifetime for {\(4d^{9}5s^2~^2D_{5/2}\)} from MCDHF/RCI is 163~ms, in reasonable agreement with the recent CI-MBPT calculation by \citet{10287TP}, which gives 192~ms. For {\(4d^{9}5s^2~^2D_{3/2}\)}, the lifetimes from the MCDHF/RCI and CI-MBPT calculations differ by a factor of three. Improved computational methodologies are urgently needed to handle transitions of this type. 
As a comment, in contrast to the regular states
in the $s$, $p$, and $d$ sequences, there is currently no experimental technique available for determining 
the lifetime of {\(4d^{9}5s^2~^2D_{5/2}\)}; its value is established entirely through theoretical calculations.

\section{Summary -- recommended data}
We have applied 
FSMRCC theory and MCDHF/RCI, two well-known many-body methods,
to compute hyperfine interaction constants, transition data and lifetimes for the $5s_{1/2}$, $6s_{1/2}$, $7s_{1/2}$, $8s_{1/2}$, $5d_{3/2,5/2}$, $6d_{3/2,5/2}$, {$4d^95s^2~^2D_{3/2,5/2}$} even and the
$5p_{1/2,3/2}$, $6p_{1/2,3/2}$, $7p_{1/2,3/2}$, $4f_{5/2,7/2}$ odd states. The results from the two methods have been compared against each other as well as with existing hyperfine and transition data from
experiment and from other calculations. We start with the hyperfine interaction constants. Whereas there is an excellent agreement between FSMRCC and MCDHF/RCI for the excited states, the hyperfine interaction constants from FSMRCC are in better agreement with experiment
for the ground and first excited states than are the MCDHF/RCI values. For this reason we recommend the FSMRCC hyperfine interaction constants. The agreement with the experimental constants is good, with a mean difference of 3\,\% for the accurately determined $5s_{1/2}, 5p_{1/2,3/2}$ states. For the {$4d^95s^2~^2D_{3/2,5/2}$} core-excited meta-stable states, the contributions from triple-excitations can not be encompassed
within the current FSMRCC implementation and for these states the values from MCDHF/RCI are closer to agreement with experiment. Turning to the weighted oscillator strengths, $gf$, the quantity most frequently used for abundance and atmospheric analysis \citep{Wahlgren}, the values from FSMRCC and MCDHF/RCI for transitions between the ordinary states of the $s$, $p$, $d$ and $f$ sequences agree very well, with a median relative differences of 2.4\,\%.  By using the 
quantitative and qualitative evaluation (QQE) approach \citep{Gaigalas.2022.p281,Kitoviene.2024.p}, as well as by looking at the 
relative changes when moving from the DHF approximation to RCCSD and further to RCCSDT,
we identified the weak $6p_{1/2,3/2} \rightarrow 5s_{1/2}$ and $7p_{1/2,3/2} \rightarrow 5s_{1/2}$ transitions as being comparatively uncertain.  The transition data, in terms of lifetimes, were 
compared with values from laser-induced fluorescence (LIF) measurements. The computed lifetimes for the higher states mainly fall within the experimental error bars. For the 
accurately known resonance transitions $5p_{1/2,3/2} \rightarrow 5s_{1/2}$,  the relative differences between the calculated lifetimes and the experimental lifetimes are 2.8\,\% and 3.8\,\%, respectively.
Based on the comparisons with the lifetimes it is not possible to single out FSMRCC or MCDHF/RCI as giving the most accurate transition data. For this reason, the recommended transition parameters are the ones obtained as the mean of the FSMRCC and MCDHF/RCI values. The uncertainties can be estimated in two ways: 
by using the relative differences of the FSMRCC and MCDHF/RCI values to assign the transitions to the NIST ASD accuracy classes or, alternatively, by using the gauge dependence 
of the MCDHF/RCI transition parameters within the QQE approach to do the assignment. We use both ways, and as the final conservative
uncertainty estimate we use the worst of the two classes. As the FSMRCC theory does not provide transition data involving the $4d^95s^2~^2D$ states, the
uncertainty estimation for the corresponding  
TOEP transitions are based entirely on the QQE approach.  The recommended
weighted oscillator strengths are displayed in Table \ref{tab:recommend} along with the estimated uncertainties. Whereas most transitions are in the higher accuracy classes, we see that the TOEP transitions mainly fall in class E. Incidentally, we note that our recommended $gf$ values for the resonance transitions are in perfect
agreement with the experimental ones from the accurate picosecond delayed-coincidence technique measurements by \cite{5944TP}, giving $gf=0.464$ and $gf = 0.952$ for
$5p_{1/2} \rightarrow 5s_{1/2}$ and $5p_{3/2} \rightarrow 5s_{1/2}$, respectively.

\begin{acknowledgements}
PJ. acknowledges support from the Swedish Research Council (VR 2023-05367).
BKS acknowledges ANRF grant no. CRG/2023/002558 and Department of Space, Government of India for financial supports. The FSMRCC calculations were carried out using the ParamVikram-1000 HPC of the Physical Research Laboratory (PRL), Ahmedabad, Gujarat, India.   
AMA acknowledges support from the Swedish Research Council (VR 2020-03940,
VR 2025-05167), and
the Crafoord Foundation via the Royal Swedish Academy of Sciences (CR
2024-0015).
\end{acknowledgements}

\bibliographystyle{aa}
\bibliography{refs}

\begin{appendix}
\onecolumn
\section{Tables}

\begin{table*}[ht!]
\begin{center}
\caption{The calculated E1 matrix elements (in a.u.) at different levels of approximation in the FSMRCC method using the length-gauge expression. The final results from the FSMRCC method are listed along with the estimated uncertainties and they are compared with the MCDHF/RCI calculations. The transitions with the largest relative difference for the matrix elements are marked with asterisks ${\ast}$, see text.}
\begin{tabular}{c|ccc c  | c|c} \hline \hline
Transition                      &  DHF   &  RCCSD & RCCSDT   & Breit  & Final  & MCDHF/RCI \\ \hline
$6s_{1/2} \rightarrow 5p_{1/2}$ & 3.403  &  2.876 & 2.786  & 0.005  & 2.791(50)  & 2.736 \\
$6s_{1/2} \rightarrow 5p_{3/2}$ & 5.073  & 4.343  & 4.212  & 0.004  & 4.216(75)  & 4.132 \\
$7s_{1/2} \rightarrow 5p_{1/2}$ & 0.704  & 0.684  & 0.688  & $\sim 0.0$ & 0.688(5)  & 0.679 \\
$7s_{1/2} \rightarrow 5p_{3/2}$ & 0.988  & 0.967  & 0.967   & $\sim 0.0$ & 0.967(5)  & 0.962 \\
$7s_{1/2} \rightarrow 6p_{1/2}$ & 8.215  & 7.468  & 7.329   & 0.008      & 7.337(76)  & 7.260 \\
$7s_{1/2} \rightarrow 6p_{3/2}$ & 12.121 & 11.100 & 10.892 & 0.006      & {10.90(10)} & 10.77\\
$8s_{1/2} \rightarrow 5p_{1/2}$ & 0.366  &  0.366 & 0.370  & $\sim 0.0$ & 0.370(4)  & 0.363\\
$8s_{1/2} \rightarrow 5p_{3/2}$ & 0.509  & 0.512  & 0.519  & $\sim 0.0$ & 0.519(5)  & 0.509\\
$8s_{1/2} \rightarrow 6p_{1/2}$ & 1.475  & 1.445  & 1.450  & $\sim 0.0$ & 1.450(6)  & 1.477\\
$8s_{1/2} \rightarrow 6p_{3/2}$ & 2.029  & 1.997  & 2.011  & 0.001      & 2.012(10)  & 2.037\\
$8s_{1/2} \rightarrow 7p_{1/2}$ & 14.864 & 13.833 & 13.615 & 0.013      & {13.63(12)} & 13.68 \\
$8s_{1/2} \rightarrow 7p_{3/2}$ & 21.836 & 20.453 & 20.133 & 0.006      & 20.139(25) & 20.130 \\
$5p_{1/2} \rightarrow 5s_{1/2}$ & 3.059  & 2.403  & 2.318  & 0.003      & 2.321(50)  & 2.188 \\
$5p_{3/2} \rightarrow 5s_{1/2}$ & 4.302  & 3.395  & 3.271  & 0.004      & 3.275(80)  & 3.098\\

$6p_{1/2} \rightarrow 5s_{1/2}$ & 0.400  & 0.183  & 0.135  & 0.001  & 0.133(25)  & 0.0816$^{\ast}$\\  
$6p_{3/2} \rightarrow 5s_{1/2}$ & 0.643  & 0.342  & 0.282  & 0.001      & 0.283(23)  & 0.195$^{\ast}$\\    
$6p_{1/2} \rightarrow 6s_{1/2}$ & 7.673  & 7.227  & 7.202  & 0.001      & 7.203(20)  & 7.127 \\
$6p_{3/2} \rightarrow 6s_{1/2}$ & 10.691 & 10.067 & 10.040   & 0.004      & 10.044(52) & 9.945 \\

$7p_{1/2} \rightarrow 5s_{1/2}$ & 0.174  & 0.056  &  0.031    & 0.001      & 0.032(10)  & 0.0102$^{\ast}$\\  
$7p_{3/2} \rightarrow 5s_{1/2}$ & 0.293  & 0.128  & 0.086    & 0.001      & 0.087(17)  & 0.0305$^{\ast}$\\   
$7p_{1/2} \rightarrow 6s_{1/2}$ & 1.036  & 0.865  & 0.825    & 0.003      & 0.828(30)  & 0.779\\
$7p_{3/2} \rightarrow 6s_{1/2}$ & 1.616  & 1.395  & 1.341    & 0.001      & 1.342(21)  & 1.269 \\
$7p_{1/2} \rightarrow 7s_{1/2}$ & 13.980 & 13.522 & 13.523   & $\sim 0.0$ & {13.52(12)} & 13.42 \\
$7p_{3/2} \rightarrow 7s_{1/2}$ & 19.405 & 18.747 & 18.762   & 0.003      & {18.76(15)} & 18.65 \\

$5d_{3/2} \rightarrow 5p_{1/2}$ & 5.700  & 4.540  & 4.399   & 0.009      & 4.408(43)  & 4.261 \\
$5d_{3/2} \rightarrow 5p_{3/2}$ & 2.654  & 2.145  & 2.075   & 0.002      & 2.077(53)  & 2.016 \\
$5d_{3/2} \rightarrow 6p_{1/2}$ & 11.980 & 11.773 & 11.722  & $-0.001$   & {11.72(10)} & 11.62\\
$5d_{3/2} \rightarrow 6p_{3/2}$ & 5.318  & 5.240  & 5.219   & $\sim 0.0$ & 5.219(86)  & 5.176\\
$5d_{3/2} \rightarrow 7p_{1/2}$ & 1.214  & 1.799  & 1.861   & $-0.008$   & 1.853(25)  & 1.870\\
$5d_{3/2} \rightarrow 7p_{3/2}$ & 0.447  & 0.689  & 0.719   & $-0.002$   &  0.717(12) & 0.730\\

$5d_{5/2} \rightarrow 5p_{3/2}$ & 7.945  & 6.416  & 6.219   & 0.009      & 6.228(16)  & 6.030 \\
$5d_{5/2} \rightarrow 6p_{3/2}$ & 16.000 & 15.769 & 15.728   & $-0.002$   & {15.73(12)} & 15.57\\
$5d_{5/2} \rightarrow 7p_{3/2}$ & 1.370  & 2.108  & 2.220   & 0.008      & 2.228(20)  & 2.226 \\

$6d_{3/2} \rightarrow 5p_{1/2}$ & 2.051 & 1.805   & 1.791 & 0.002      & 1.793(51)  & 1.700 \\
$6d_{3/2} \rightarrow 5p_{3/2}$ & 0.923  & 0.824  & 0.808  & $\sim 0.0$ & 0.808(28)  & 0.778 \\
$6d_{3/2} \rightarrow 6p_{1/2}$ & 9.332  & 7.657  & 7.418  & 0.018      & 7.436(69)  & 7.310 \\
$6d_{3/2} \rightarrow 6p_{3/2}$ & 4.406  & 3.671  & 3.554   & 0.005      & 3.559(47)  & 3.502\\
$6d_{3/2} \rightarrow 7p_{1/2}$ & 24.483 & 24.176 & 24.135   & $-0.001$   & 24.135(67) & 24.122 \\
$6d_{3/2} \rightarrow 7p_{3/2}$ & 10.886 & 10.777 & 10.766   & $\sim 0.0$ & 10.766(41) & 11.75 \\
      
$6d_{5/2} \rightarrow 5p_{3/2}$ & 2.773  & 2.474  & 2.406  & $\sim 0.0$ &  2.406(22)  & 2.333 \\
$6d_{5/2} \rightarrow 6p_{3/2}$ & 13.156 & 10.936 & 10.616   & 0.018      & {10.63(11)} & 10.44 \\
$6d_{5/2} \rightarrow 7p_{3/2}$ & 32.723 & 32.407 & 32.340  & 0.003      & 32.343(22) & 32.322 \\

$4f_{5/2} \rightarrow 5d_{3/2}$ & 16.563 & 15.704 & 15.556   & 0.002      & 15.558(30) & 15.321 \\
$4f_{5/2} \rightarrow 6d_{3/2}$ & 23.873 & 24.109 & 24.130   & $-0.001$   & 24.129(38) & 24.510 \\
$4f_{5/2} \rightarrow 5d_{5/2}$ & 4.440  & 4.213  & 4.283   & $\sim 0.0$ & 4.283(46)  & 4.109 \\
$4f_{7/2} \rightarrow 5d_{5/2}$ & 19.855 & 18.842 & 18.706  & $\sim 0.0$ & 18.706(45) & 18.375 \\
$4f_{5/2} \rightarrow 6d_{5/2}$ & 6.372  & 6.434  &  6.444  & $\sim 0.0$ & 6.444(55)  & 6.544\\
$4f_{7/2} \rightarrow 6d_{5/2}$ & 28.495 & 28.772 & 28.808 & 0.001      & 28.809(86) & 29.263\\
\hline 
\end{tabular}
\label{tab_E1}
\end{center}
\end{table*}

   \begin{table}
    \begin{center}
          \caption{E1 transition rates \(A\) (in length gauge) in \(\mathrm{s}^{-1}\) from FSMRCC and MCDHF/RCI calculations, along with uncertainty estimates for the latter based on the QQE approach. Here, \(4d^{-1}\) denotes the core-excited configuration {\(4d^95s^2\)}.} 
     \footnotesize
     \begin{tabular}{cccl} \hline \hline
      Transition                   &  FSMRCC     & MCDHF/RCI &  unc.       \\ \hline
      $6s_{1/2} \rightarrow 5p_{1/2}$ & 1.735+07 & 1.877+07 & A+           \\
      $6s_{1/2} \rightarrow 5p_{3/2}$ & 3.177+07 & 3.414+07 & A+      \\ \hline
                                                 
      $7s_{1/2} \rightarrow 5p_{1/2}$ & 5.343+06 & 5.685+06 & AA      \\
      $7s_{1/2} \rightarrow 5p_{3/2}$ & 9.303+06 & 1.002+07 & AA      \\
      $7s_{1/2} \rightarrow 6p_{1/2}$ & 2.522+06 & 2.565+06 & A+      \\
      $7s_{1/2} \rightarrow 6p_{3/2}$ & 4.671+06 & 4.712+06 & AA      \\  \hline
                                                 
      $8s_{1/2} \rightarrow 5p_{1/2}$ & 2.446+06 & 2.555+06 & A+      \\
      $8s_{1/2} \rightarrow 5p_{3/2}$ & 4.320+06 & 4.492+06 & A       \\
      $8s_{1/2} \rightarrow 6p_{1/2}$ & 8.231+05 & 8.859+05 & B       \\
      $8s_{1/2} \rightarrow 6p_{3/2}$ & 1.456+06 & 1.544+06 & B+      \\
      $8s_{1/2} \rightarrow 7p_{1/2}$ & 6.853+05 & 6.922+05 & B+      \\
      $8s_{1/2} \rightarrow 7p_{3/2}$ & 1.279+06 & 1.270+06 & B+      \\  \hline
                                                 
      $5p_{1/2} \rightarrow 5s_{1/2}$ & 1.409+08 & 1.390+08 & C+      \\  \hline
                                                 
      $5p_{3/2} \rightarrow 5s_{1/2}$ & 1.537+08 & 1.531+08 & C+      \\
 $5p_{3/2} \rightarrow 4d^{-1}_{5/2}$ & --       & 3.57+00  & E       \\  \hline
                                                 
      $6p_{1/2} \rightarrow 5s_{1/2}$ & 2.019+06 & 8.426+05 & E*      \\ 
 $6p_{1/2} \rightarrow 4d^{-1}_{3/2}$ & --       & 4.137+04 & E       \\
      $6p_{1/2} \rightarrow 6s_{1/2}$ & 9.947+06 & 1.026+07 & A+      \\  \hline
                                                 
      $6p_{3/2} \rightarrow 5s_{1/2}$ & 4.628+06 & 2.446+06 & E*      \\ 
 $6p_{3/2} \rightarrow 4d^{-1}_{5/2}$ & --       & 1.668+05 & E       \\
 $6p_{3/2} \rightarrow 4d^{-1}_{3/2}$ & --       & 5.026+03 & E       \\
      $6p_{3/2} \rightarrow 6s_{1/2}$ & 1.074+07 & 1.112+07 & A+      \\  \hline
                                                 
      $7p_{1/2} \rightarrow 5s_{1/2}$ & 1.637+05 & 1.834+04 & E*      \\ 
 $7p_{1/2} \rightarrow 4d^{-1}_{3/2}$ & --       & 7.304+04 & E       \\
      $7p_{1/2} \rightarrow 6s_{1/2}$ & 1.052+06 & 9.800+05 & B+      \\
      $7p_{1/2} \rightarrow 5d_{3/2}$ & 5.170+05 & 5.426+05 & A       \\
      $7p_{1/2} \rightarrow 7s_{1/2}$ & 1.852+06 & 1.938+06 & AA      \\  \hline
                                                 
      $7p_{3/2} \rightarrow 5s_{1/2}$ & 6.078+05 & 8.22+04  & E*      \\ 
 $7p_{3/2} \rightarrow 4d^{-1}_{5/2}$ & --       & 1.971+05 & E       \\
 $7p_{3/2} \rightarrow 4d^{-1}_{3/2}$ & --       & 8.511+03 & E       \\
      $7p_{3/2} \rightarrow 6s_{1/2}$ & 1.411+06 & 1.325+06 & B+      \\
      $7p_{3/2} \rightarrow 5d_{3/2}$ & 4.049+04 & 4.330+04 & A       \\
      $7p_{3/2} \rightarrow 5d_{5/2}$ & 3.865+05 & 3.982+05 & B+      \\
      $7p_{3/2} \rightarrow 7s_{1/2}$ & 1.989+06 & 2.089+06 & AA      \\  \hline

 $4d^{-1}_{3/2} \rightarrow 5p_{1/2}$ & --       & 1.637+04 & E     \\
 $4d^{-1}_{3/2} \rightarrow 5p_{3/2}$ & --       & 1.729+03 & E     \\  \hline

      $5d_{3/2} \rightarrow 5p_{1/2}$ & 6.957+07 & 7.194+07 & B+    \\
      $5d_{3/2} \rightarrow 5p_{3/2}$ & 1.333+07 & 1.383+07 & B+    \\
      $5d_{3/2} \rightarrow 6p_{1/2}$ & 6.197+03 & 7.732+03 & B+    \\
      $5d_{3/2} \rightarrow 6p_{3/2}$ & 1.983+02 & 2.707+02 & B     \\  \hline
                                                 
      $5d_{5/2} \rightarrow 5p_{3/2}$ & 8.016+07 & 8.273+07 & B+    \\
      $5d_{5/2} \rightarrow 6p_{3/2}$ & 1.526+03 & 1.977+03 & B     \\  \hline
                                                 
      $6d_{3/2} \rightarrow 5p_{1/2}$ & 2.439+07 & 2.386+07 & B+    \\
      $6d_{3/2} \rightarrow 5p_{3/2}$ & 4.419+06 & 4.438+06 & B+    \\
      $6d_{3/2} \rightarrow 6p_{1/2}$ & 5.769+06 & 5.818+06 & AA    \\
      $6d_{3/2} \rightarrow 6p_{3/2}$ & 1.190+06 & 1.199+06 & AA    \\
      $6d_{3/2} \rightarrow 7p_{1/2}$ & 1.256+03 & 1.148+03 & B+    \\
      $6d_{3/2} \rightarrow 7p_{3/2}$ & 3.238+01 & 2.465+01 & C+    \\  \hline
                                                 
      $6d_{5/2} \rightarrow 5p_{3/2}$ & 2.616+07 & 2.665+07 & B+    \\
      $6d_{5/2} \rightarrow 6p_{3/2}$ & 7.119+06 & 7.137+06 & AA    \\
      $6d_{5/2} \rightarrow 7p_{3/2}$ & 2.792+02 & 2.103+02 & C+    \\  \hline
                                                 
 $4f_{5/2} \rightarrow 4d^{-1}_{3/2}$ &  --      & 1.533+04 & A+    \\
 $4f_{5/2} \rightarrow 4d^{-1}_{5/2}$ &  --      & 1.146+03 & B+    \\
      $4f_{5/2} \rightarrow 5d_{3/2}$ & 1.331+07 & 1.354+07 & AA    \\ 
      $4f_{5/2} \rightarrow 5d_{5/2}$ & 9.975+05 & 9.646+05 & AA    \\  \hline
                                                 
 $4f_{7/2} \rightarrow 4d^{-1}_{5/2}$ & --       & 1.688+04 & A     \\
      $4f_{7/2} \rightarrow 5d_{5/2}$ & 1.427+07 & 1.447+07 & AA    \\ \hline 
     \end{tabular}
     \label{tab:transrate}
    \end{center}
 \end{table}

    \begin{table}
     \begin{center}
          \caption{Weighted oscillator strengths \(gf\) in length gauge from FSMRCC and MCDHF/RCI calculations, along with uncertainty estimates for the latter based on the QQE approach. Here, \(4d^{-1}\) denotes the core-excited configuration {\(4d^95s^2\)}. FMP refers to values reported by \citet{8781TP} based on the Fues model potential.} 
     \footnotesize
     \begin{tabular}{cc|ll|l} \hline \hline
      Transition                      &  FSMRCC        & MCDHF/RCI &  unc. & FMP    \\ \hline
      $6s_{1/2} \rightarrow 5p_{1/2}$ &  0.308      & 0.308     & A+    & 0.332 \\
      $6s_{1/2} \rightarrow 5p_{3/2}$ &  0.652      & 0.651     & A+    & 0.684 \\ \hline  
                                                 
      $7s_{1/2} \rightarrow 5p_{1/2}$ &  0.0321     & 0.0322    & AA    & 0.0298 \\
      $7s_{1/2} \rightarrow 5p_{3/2}$ &  0.0608     & 0.0619    & AA    & 0.0552 \\
      $7s_{1/2} \rightarrow 6p_{1/2}$ &  0.587      & 0.582     & A+    & 0.644\\
      $7s_{1/2} \rightarrow 6p_{3/2}$ &  1.222      & 1.206     & AA    & 1.308 \\  \hline
                                                 
      $8s_{1/2} \rightarrow 5p_{1/2}$ &  0.0108     & 0.0107    & A+    & 0.00960 \\
      $8s_{1/2} \rightarrow 5p_{3/2}$ &  0.0205     & 0.0203    & A     & 0.0176 \\
      $8s_{1/2} \rightarrow 6p_{1/2}$ &  0.0465     & 0.0489    & B     & 0.0466 \\
      $8s_{1/2} \rightarrow 6p_{3/2}$ &  0.08706    & 0.0903    & B+    & 0.0848 \\
      $8s_{1/2} \rightarrow 7p_{1/2}$ &  0.869      & 0.876     & B+    & 0.934\\
      $8s_{1/2} \rightarrow 7p_{3/2}$ &  1.799      & 1.794     & B+    & 1.896\\  \hline
                                                 
      $5p_{1/2} \rightarrow 5s_{1/2}$ &  0.484      & 0.445     & C+    & 0.558 \\  \hline
                                                 
      $5p_{3/2} \rightarrow 5s_{1/2}$ &  0.993      & 0.921     & C+    &  1.14\\
 $5p_{3/2} \rightarrow 4d^{-1}_{5/2}$ & --          & 0.000314  & E     & -- \\  \hline
                                                 
      $6p_{1/2} \rightarrow 5s_{1/2}$ &  0.00259    & 0.00101   & E*    & 0.0222\\ 
 $6p_{1/2} \rightarrow 4d^{-1}_{3/2}$ & --          & 0.000624  & E     & -- \\
      $6p_{1/2} \rightarrow 6s_{1/2}$ &  0.905      & 0.902     & A+    & 0.786 \\  \hline
                                                 
      $6p_{3/2} \rightarrow 5s_{1/2}$ &  0.0118     & 0.0582    & E*   & 0.0530 \\ 
 $6p_{3/2} \rightarrow 4d^{-1}_{5/2}$ & --          & 0.00281   & E     & --\\
 $6p_{3/2} \rightarrow 4d^{-1}_{3/2}$ & --          & 0.000147  & E     & -- \\
      $6p_{3/2} \rightarrow 6s_{1/2}$ &  1.822      & 1.819     & A+    & 1.584 \\  \hline
                                                 
      $7p_{1/2} \rightarrow 5s_{1/2}$ &  0.000168   & 0.0000177 & E*    & 0.00568 \\ 
 $7p_{1/2} \rightarrow 4d^{-1}_{3/2}$ & --          & 0.000551  & E     & -- \\
      $7p_{1/2} \rightarrow 6s_{1/2}$ &  0.0239     & 0.0215    & B+    & 0.0308 \\
      $7p_{1/2} \rightarrow 5d_{3/2}$ &  0.0552     & 0.0568    & A     & --\\
      $7p_{1/2} \rightarrow 7s_{1/2}$ & 1.1966      & 1.202     & AA    & 1.086 \\  \hline
                                                 
      $7p_{3/2} \rightarrow 5s_{1/2}$ &  0.00124    & 0.00158   & E*    & 0.0140 \\ 
 $7p_{3/2} \rightarrow 4d^{-1}_{5/2}$ & --          & 0.00196   & E     & --\\
 $7p_{3/2} \rightarrow 4d^{-1}_{3/2}$ & --          & 0.000127  & E     & --\\
      $7p_{3/2} \rightarrow 6s_{1/2}$ &  0.0633     & 0.0575    & B+    & 0.0734 \\
      $7p_{3/2} \rightarrow 5d_{3/2}$ &  0.00840    & 0.00880   & A     & --\\
      $7p_{3/2} \rightarrow 5d_{5/2}$ &  0.0808     & 0.0814    & B+    & --\\
      $7p_{3/2} \rightarrow 7s_{1/2}$ &  2.3906     & 2.409     & AA    & 2.180 \\  \hline

 $4d^{-1}_{3/2} \rightarrow 5p_{1/2}$ & --       & 0.00353   & E     & --\\
 $4d^{-1}_{3/2} \rightarrow 5p_{3/2}$ & --       & 0.000563  & E     & --\\  \hline

      $5d_{3/2} \rightarrow 5p_{1/2}$ &  1.133      & 1.095     & B+    & 1.194 \\
      $5d_{3/2} \rightarrow 5p_{3/2}$ &  0.239      & 0.233     & B+    & 0.248 \\
      $5d_{3/2} \rightarrow 6p_{1/2}$ &  0.186      & 0.198     & B+    & 0.182\\
      $5d_{3/2} \rightarrow 6p_{3/2}$ &  0.0201     & 0.0201    & B     & 0.0197 \\  \hline
                                                 
      $5d_{5/2} \rightarrow 5p_{3/2}$ &  2.155      & 2.086     & B+    & 2.232\\
      $5d_{5/2} \rightarrow 6p_{3/2}$ &  0.198      & 0.213     & B     & -- \\  \hline
                                                 
      $6d_{3/2} \rightarrow 5p_{1/2}$ &  0.241      & 0.223     & B+    & 0.264 \\
      $6d_{3/2} \rightarrow 5p_{3/2}$ &  0.0471     & 0.0448    & B+    & 0.0524 \\
      $6d_{3/2} \rightarrow 6p_{1/2}$ &  0.992      & 0.972     & AA    & 1.016\\
      $6d_{3/2} \rightarrow 6p_{3/2}$ &  0.219      & 0.215     & AA    & 0.220\\
      $6d_{3/2} \rightarrow 7p_{1/2}$ &  0.287      & 0.278     & B+    & 0.284\\
      $6d_{3/2} \rightarrow 7p_{3/2}$ &  0.0289     & 0.0263    & C+    & 0.0286\\  \hline
                                                 
      $6d_{5/2} \rightarrow 5p_{3/2}$ &  0.417      & 0.403     & B+    & 0.472 \\
      $6d_{5/2} \rightarrow 6p_{3/2}$ &  1.962      & 1.917     & AA    & --\\
      $6d_{5/2} \rightarrow 7p_{3/2}$ &  0.294      & 0.267     & C+    & --\\  \hline
                                                 
 $4f_{5/2} \rightarrow 4d^{-1}_{3/2}$ &  --        & 0.000341  & A+    & -- \\
 $4f_{5/2} \rightarrow 4d^{-1}_{5/2}$ &  --        & 0.0000169 & B+    & -- \\
      $4f_{5/2} \rightarrow 5d_{3/2}$ &  4.015     & 3.956     & AA    & 4.038 \\
      $4f_{5/2} \rightarrow 5d_{5/2}$ &  0.303     & 0.284     & AA    & 0.289\\   \hline
                                                 
 $4f_{7/2} \rightarrow 4d^{-1}_{5/2}$ & --         & 0.000332  & A     & -- \\
      $4f_{7/2} \rightarrow 5d_{5/2}$ &  5.782     & 5.672     & AA    & 5.797 \\ \hline 
     \end{tabular}
     \label{tab:rategf}
     \end{center}
 \end{table}

   \begin{table*}
  \begin{center}
     \caption{Lifetimes from MCDHF/RCI and FSMRCC calculations compared with experimental values and other theories. Error bars to the FSMRCC values are estimated using the uncertainties of the calculated E1 matrix elements.}
     \begin{tabular}{lrrrrrr} \hline \hline
      State  &  MCDHF/RCI   & FSMRCC      & FSMRCC$^{{(a)}}$   &  MBPT$^{{(b)}}$   & CI-MBPT$^{{(c)}}$ &  Experiment \\ \hline
      $4d^{10}5p~^2P_{1/2}$  & 7.20 ns  & 7.10(30) ns  & 6.718 ns & 7.62 ns    & 6.6 ns  & 7.41(0.04) ns$^{{(d)}}$ \\
      $4d^{10}5p~^2P_{3/2}$    & 6.53 ns  & 6.51(30) ns  & 6.193 ns & 6.97 ns    & 6.1 ns  & 6.79(0.03) ns$^{{(d)}}$\\
      $4d^{9}5s^2~^2D_{5/2}$   & 163 ms  &  &   &   & 192 ms   & \\
      $4d^{9}5s^2~^2D_{3/2}$   & 55.2 $\mu$s  &    &    &   & 174 $\mu$s & \\
      $4d^{10}6s~^2S_{1/2}$    & 18.9 ns  & 20.36(71) ns &   &   &   & 22(3) ns$^{{(e)}}$ \\
      $4d^{10}6p~^2P_{1/2}$   & 89.7 ns  & 99(15) ns  &    &   &   & 95(6) ns$^{{(f)}}$ \\
      $4d^{10}6p~^2P_{3/2}$    & 72.8 ns  & 65(5) ns  &    &   &  & 78(5) ns$^{{(f)}}$ \\
      $4d^{10}5d~^2D_{3/2}$    & 11.7 ns   & 12.06(29) ns  &    &   &    & 11(3) ns$^{{(e)}}$\\
      $4d^{10}5d~^2D_{5/2}$    & 12.1 ns  & 12.48(30) ns  &     &   &    & \\
      $4d^{10}7s~^2S_{1/2}$    & 43.5 ns  & 45.79(65) ns  & & & & 53(5) ns$^{{(e)}}$\\
      $4d^{10}7p~^2P_{1/2}$    & 282 ns   & 279(18) ns  &  &  &   & 285(25) ns$^{{(g)}}$ \\
      $4d^{10}7p~^2P_{3/2}$    & 241 ns       & 225(16) ns  &          &            &            & 255(20) ns$^{{(g)}}$\\
      $4d^{10}6d~^2D_{3/2}$    & 28.3 ns      & 28(1) ns  &          &            &            & 26(4) ns$^{{(e)}}$\\ 
      $4d^{10}4f~^2F_{5/2}$    & 68.9 ns      & 69.89(36) ns  &          &            &            & \\
      $4d^{10}4f~^2F_{7/2}$    & 69.0 ns      & 70.07(34) ns  &          &            &            & \\
      $4d^{10}6d~^2D_{5/2}$    & 29.6 ns      & 30.05(56) ns  &          &            &            & \\
      $4d^{10}8s~^2S_{1/2}$    & 87.4 ns      & 91(2) ns  &          &            &            & 98(10) ns$^{{(e)}}$\\ \hline 
     \end{tabular}

\tablefoot{
\tablefoottext{a}{\cite{10.1063/5.0128225}};
\tablefoottext{b}{\cite{PhysRevA.68.062505}};
\tablefoottext{c}{\cite{10287TP}};
\tablefoottext{d}{\cite{5944TP}};
\tablefoottext{e}{\cite{5681TP}};
\tablefoottext{f}{\cite{5647TP}};
\tablefoottext{g}{\cite{5835TP}}}
 \label{tab:lifetime}
    \end{center}
 \end{table*}

    \begin{table*}
    \begin{center}
          \caption{Recommended weighted oscillator strengths, $gf$, taken as the mean of the FSMRCC and MCDHF/RCI values, along with estimated uncertainties. For transitions
          involving $4d^95s^2$, here abbreviated as \(4d^{-1}\), the values are from MCDHF/RCI. The uncertainties in column three in terms of the NIST ASD accuracy classes are based the relative differences between the FSMRCC and MCDHF/RCI values and the uncertainties in column four are based on the QQE approach. The final uncertainties, column five, 
          are the worst of column 3 and 4. For transitions involving $4d^95s^2$ the QQE approach is used for the uncertainty estimation.} 
     \footnotesize
     \begin{tabular}{cc|llll} \hline \hline
      Transition                      &  mean    & unc. rel. diff.  & unc. QQE & unc. final  \\ \hline
      $6s_{1/2} \rightarrow 5p_{1/2}$ & 0.308    & AA   & A+  &  A+ \\
      $6s_{1/2} \rightarrow 5p_{3/2}$ & 0.651    & AA   & A+  &  A+ \\ \hline
                                                 
      $7s_{1/2} \rightarrow 5p_{1/2}$ & 0.0321   & AA   & AA  & AA \\
      $7s_{1/2} \rightarrow 5p_{3/2}$ & 0.0613   & A+   & AA  & A+   \\
      $7s_{1/2} \rightarrow 6p_{1/2}$ & 0.584    & AA   & A+  & A+ \\
      $7s_{1/2} \rightarrow 6p_{3/2}$ & 1.214    & A+   & AA  & A+  \\  \hline
                                                 
      $8s_{1/2} \rightarrow 5p_{1/2}$ & 0.0107   & AA   & A+  & A+  \\
      $8s_{1/2} \rightarrow 5p_{3/2}$ & 0.0204   & AA   & A   & A \\
      $8s_{1/2} \rightarrow 6p_{1/2}$ & 0.0477   & B+   & B   & B   \\
      $8s_{1/2} \rightarrow 6p_{3/2}$ & 0.0887   & B+   & B+  & B+ \\
      $8s_{1/2} \rightarrow 7p_{1/2}$ & 0.872    & AA   & B+  & B+  \\
      $8s_{1/2} \rightarrow 7p_{3/2}$ & 1.796    & AA   & B+  & B+  \\  \hline
                                                 
      $5p_{1/2} \rightarrow 5s_{1/2}$ & 0.464    & B    & C+ & C+  \\  \hline
                                                 
      $5p_{3/2} \rightarrow 5s_{1/2}$ & 0.957    & B   & C+ & C+  \\
 $5p_{3/2} \rightarrow 4d^{-1}_{5/2}$ & 0.000314 & --  & E & E\\  \hline
                                                 
      $6p_{1/2} \rightarrow 5s_{1/2}$ & 0.00180  & E  & E & E \\ 
 $6p_{1/2} \rightarrow 4d^{-1}_{3/2}$ & 0.000624 & --  & E  & E\\
      $6p_{1/2} \rightarrow 6s_{1/2}$ & 0.903    & AA   & A+ & A+  \\  \hline
                                                 
      $6p_{3/2} \rightarrow 5s_{1/2}$ & 0.0350    &  E  & E & E\\ 
 $6p_{3/2} \rightarrow 4d^{-1}_{5/2}$ & 0.00281  &  -- & E  & E  \\
 $6p_{3/2} \rightarrow 4d^{-1}_{3/2}$ & 0.000147 &  -- & E  & E \\
      $6p_{3/2} \rightarrow 6s_{1/2}$ & 1.820    & AA  & A+ & A+  \\  \hline
                                                 
      $7p_{1/2} \rightarrow 5s_{1/2}$ & 0.0000928   &  E  & E & E \\ 
 $7p_{1/2} \rightarrow 4d^{-1}_{3/2}$ & 0.000551 & --   & E & E\\
      $7p_{1/2} \rightarrow 6s_{1/2}$ & 0.0227   & B   & B+ & B \\
      $7p_{1/2} \rightarrow 5d_{3/2}$ & 0.0560   & A    & A     & A\\
      $7p_{1/2} \rightarrow 7s_{1/2}$ & 1.199    & AA   & AA    & AA   \\  \hline
                                                 
      $7p_{3/2} \rightarrow 5s_{1/2}$ & 0.00141  & C    & E & E \\ 
 $7p_{3/2} \rightarrow 4d^{-1}_{5/2}$ & 0.00196  & --   & E & E  \\
 $7p_{3/2} \rightarrow 4d^{-1}_{3/2}$ & 0.000127 & --   & E & E\\
      $7p_{3/2} \rightarrow 6s_{1/2}$ & 0.0604   & B    & B+ & B  \\
      $7p_{3/2} \rightarrow 5d_{3/2}$ & 0.00860  & B+   & A  & B+\\
      $7p_{3/2} \rightarrow 5d_{5/2}$ & 0.0811   & AA   & AA & AA \\
      $7p_{3/2} \rightarrow 7s_{1/2}$ & 2.400    & AA   & AA & AA \\  \hline

 $4d^{-1}_{3/2} \rightarrow 5p_{1/2}$ & 0.00353  &  -- & E & E \\
 $4d^{-1}_{3/2} \rightarrow 5p_{3/2}$ & 0.000563 &  -- & E & E \\  \hline

      $5d_{3/2} \rightarrow 5p_{1/2}$ & 1.114    & B+  & B+ & B+    \\
      $5d_{3/2} \rightarrow 5p_{3/2}$ & 0.236    & A   & B+ & B+    \\
      $5d_{3/2} \rightarrow 6p_{1/2}$ & 0.192    & B+  & B+ & B+    \\
      $5d_{3/2} \rightarrow 6p_{3/2}$ & 0.0201   & AA  & B  & B   \\  \hline
                                                 
      $5d_{5/2} \rightarrow 5p_{3/2}$ & 2.120    & B+  & B+ & B+  \\
      $5d_{5/2} \rightarrow 6p_{3/2}$ & 0.205    & B+  & B  & B  \\  \hline
                                                 
      $6d_{3/2} \rightarrow 5p_{1/2}$ & 0.232    & B   & B+ & B \\
      $6d_{3/2} \rightarrow 5p_{3/2}$ & 0.0459   & B+  & B+ & B+ \\
      $6d_{3/2} \rightarrow 6p_{1/2}$ & 0.982    & A   & AA & A   \\
      $6d_{3/2} \rightarrow 6p_{3/2}$ & 0.217    & A+  & AA & A+   \\
      $6d_{3/2} \rightarrow 7p_{1/2}$ & 0.282    & B+  & B+ & B+  \\ 
      $6d_{3/2} \rightarrow 7p_{3/2}$ & 0.0276   & B   & C+ & C+ \\  \hline
                                                 
      $6d_{5/2} \rightarrow 5p_{3/2}$ & 0.410    & B+  & B+ & B+  \\
      $6d_{5/2} \rightarrow 6p_{3/2}$ & 1.939    & A   & AA & A  \\
      $6d_{5/2} \rightarrow 7p_{3/2}$ & 0.280    & B   & C+ & C+   \\  \hline
                                                 
 $4f_{5/2} \rightarrow 4d^{-1}_{3/2}$ &  0.000341  & -- & A+ & A+ \\
 $4f_{5/2} \rightarrow 4d^{-1}_{5/2}$ &  0.0000169 & -- & B+ & B+ \\
      $4f_{5/2} \rightarrow 5d_{3/2}$ & 3.985    & A+   & AA & A+    \\
      $4f_{5/2} \rightarrow 5d_{5/2}$ & 0.293    & B+   & AA & B+  \\   \hline
                                                 
 $4f_{7/2} \rightarrow 4d^{-1}_{5/2}$ & 0.000332       & -- & A & A\\
      $4f_{7/2} \rightarrow 5d_{5/2}$ & 5.727    & A+ & AA & A+   \\ \hline 
     \end{tabular}
     \label{tab:recommend}
     \end{center}
 \end{table*}

\end{appendix}

\end{document}